\documentclass[printer]{aa}
\usepackage{longtable} 
\usepackage{supertabular}
\usepackage{graphicx}
\usepackage{amssymb}
\usepackage{amsmath}
\usepackage{units}
\usepackage{multirow}
\usepackage{version}
\usepackage{color}
\usepackage{float}
\usepackage{natbib}

\reversemarginpar

\def\be{\begin{equation}}
\def\ee{\end{equation}}
\def\bi{\begin{itemize}}
\def\ei{\end{itemize}}
\newcommand{\Archeops}{{\sc Archeops}}
\newcommand{\Wmap}{{\sc Wmap}}
\newcommand{\Planck}{{\sc Planck}}
\newcommand{\Iris}{{\sc Iris}}
\newcommand{\Firas}{{\sc Firas}}
\newcommand{\Dirbe}{{\sc Dirbe}}
\newcommand{\Iras}{{\sc Iras}}

\begin{document}

\title{Expected constraints on the Galactic magnetic field using PLANCK data} 

 \author{ L.~Fauvet~\inst{1,2}\and J.~F.~Mac\'{\i}as-P\'erez~\inst{2}
   \and T. R. Jaffe~\inst{5} \and A. J. Banday~\inst{5} \and
   F.X. D\'esert~\inst{2,3,4} \and D. Santos~\inst{2}}

\institute{European Space Agency (ESA), Research and Scientific
  Support Dpt., Astrophysics Division, Keplerlaan 1, 2201AZ Noordwijk,
  The Netherlands \and LPSC, Universit\'e Joseph Fourier Grenoble 1, CNRS/IN2P3,
  Institut National Polytechnique de Grenoble, 53 avenue des Martyrs,
  38026 Grenoble cedex, France \and IPAG, Institut de Plan\'etologie
  et d'Astrophysique de Grenoble, UJF-Grenoble 1 CNRS/INSU, UMR 524,
  Grenoble, F-38041, France \and Institut N\'eel, 25 rue des Martyrs,
  BP 166, 38042 Grenoble cedex 9, France \and Institut de Recherche en
Astrophysique et Plan\'etologie, Universit\'e de Toulouse (UPS-OMP),
CNRS, UMR 5277, 9 Av. du colonel Roche, 31028 Toulouse, France}

 \abstract{}{We explore in this paper the ability to constrain the Galactic magnetic field
intensity and spatial distribution with the incoming data from the
\Planck\ satellite experiment.}{We perform realistic simulations of the
\Planck\ observations at the polarized frequency bands from 30 to 353
GHz for two all--sky surveys as expected for the nominal
mission. These simulations include CMB, synchrotron and thermal dust
Galactic emissions and instrumental noise. (Note that systematic
effects are not considered in this paper). For the synchrotron and
thermal dust Galactic emissions we use a coherent 3D model of the
Galaxy describing its mater density and the magnetic field direction and
intensity.
We first simulate the synchrotron and dust emissions at 408~MHz and 545~GHz, respectively,
and then we extrapolate them to the \Planck\ frequency bands.}
{We perform a likelihood analysis to compare the simulated
data to a set of models obtained by varying the pitch angle of the regular magnetic field spatial distribution,
the relative amplitude of the turbulent magnetic field,
the radial scale of the electron and dust grain distributions, and the extrapolation spectral indices for the synchrotron
and thermal dust emissions. We are able to set tight constraints on
all the parameters considered. We have also found that the observed spatial variations of the
synchrotron and thermal dust spectral indices should not affect our ability to recover the other parameters of the model.}{From this, we conclude that the
\Planck\ satellite experiment can precisely measure the main properties of the Galactic magnetic field.
An accurate reconstruction of the matter distribution would require on
the one hand an improved modelling of the
ISM and on the other hand to use extra data sets like rotation measurements of pulsars.} 
\keywords{ISM: general -- ISM: clouds -- Methods: data analysis -- Cosmology: observations -- Submillimeter  }

\date{\today}

\titlerunning{Galactic polarized foreground for PLANCK}
\maketitle






\section{Introduction}

\indent The \Planck\ satellite~\citep{early1, tauber}, currently in flight, will provide
measurements of the CMB anisotropies both in temperature and
polarization over the full-sky at unprecedented accuracy. It covers a large range of frequencies from 30
to 857 GHz and therefore is able to give a measurement of the
foreground emissions. \Planck\
makes observations of the sky with a combined sensitivity of $\Delta T/T_{CMB} \sim 2 \times
10^{-6}$~\citep{bluebook} and an angular resolution from 33 to 5
arcmin~\citep{bluebook}. In particular, because of
its 7 polarized channels it will for the first time allow the simultaneous precise measurement of the
main sources of polarized Galactic emissions: synchrotron and thermal dust.\\

 \indent Using the \Wmap\ (\emph{Wilkinson Microwave Anisotropy
   Probe}), \cite{page2007} have shown that the synchrotron emission
 is highly polarized, up to 70\% between 23 and 94 GHz. Furthermore,
 \cite{benoit2004a, ponthieu2005} have observed significantly
 polarized thermal dust emission, up to 15 \% in the 353 GHz
 \Archeops\ channel. The free-free emission is not intrinsically
 polarized and the anomalous dust-correlated emission is weakly
 polarized at $3^{+1.3}_{-1.9}$~\citep{battistelli2006}. At the \Planck\
 frequency bands the polarized contribution from compact and point
 sources is expected to be weak for both radio~\citep{nolta2009} and
 dusty~\citep{desert1998} sources. The frequency and spatial
 distributions of the polarized diffuse Galactic emissions are not
 currently well known and the only available informations come from
 microwave and submillimeter observations.\\

\indent The Galactic synchrotron emission originate from relativistic
electrons spiraling along magnetic field lines. The direction of
polarization is orthogonal both to the line-of-sight and to the field
lines.\\

\indent The synchrotron emission contributes principally to diffuse
emission at both radio and microwave observation frequencies. Its
spectral energy distribution (SED) is not known with accuracy however
it is assumed to be well reproduced by a power law in antenna
temperature $T_{\nu} \alpha \nu^{\beta_s}$ with a spectral index
ranging form -2.7 to -3.4~\citep{lawson, reich88, reich04, kogut,
  gold, fauvet} . The Galactic synchrotron emission has been well traced by the
Leiden survey between 408 MHz and 1.4 GHz~\citep{brouw, wolleben}, the
Parkes survey at 2.4 GHz~\citep{duncan1999} and the MGLS (
\emph{Medium Galactic Latitude Survey}) at 1.4
GHz~\citep{uyaniker}. At the microwave frequencies it has been mapped
by \Wmap, see e.g.~\cite{hinshaw, page2007}.\\

\indent Thermal dust emission arises from dust grains in the interstellar
medium (ISM) with typical sizes $\simeq$ 0.25 $\mu$m that are heated by stellar
radiation~\citep{desert1998}. This emission can be partially
  polarized as prolate dust grains align with their long axis
  perpendicular to the magnetic field \citep{davis}. The dust emission efficiency
is greatest along the long axis, leading to partial linear polarization
perpendicular to the magnetic field. The fractional polarization depends
on the grain size distribution and is typically a few percent at
millimeter wavelengths~\citep{hildebrand1999, vaillancourt,
  fauvet}. The thermal dust emission in intensity has already been
well measured by \Iras\ form 5 to 100 $\mu$m~\citep{neugebauer} and
COBE/FIRAS which provided the first polarized observation at high
frequencies. Currently, Planck HFI~\citep{cote} is measuring this emission
in intensity \citep{early19, early21, early24, early25}.\\ 

\indent Based on the physical characteristics of the synchrotron
emission, \cite{page2007} proposed a 3D model of the Galaxy including
the distribution of relativistic electrons and the spatial
distribution of the Galactic magnetic field. Independently,
\cite{han2004, han2006} used a 3D model of the free electrons density
in the Galaxy~\citep{cordes} and a model of the Galactic magnetic
field including regular and turbulent components to explain the
observed rotation measurements toward known pulsars. Based on these
works~\cite{sun} performed a combined analysis of the rotation
measurement of pulsars and of the polarized \Wmap\ data. This work has
been extended by~\cite{jaffe} to study the Galactic plane and
\cite{jansson}  for the full sky. Recently \cite{fauvet} proposed for
the first time a coherent model of the synchrotron and thermal dust
Galactic emissions using the \Wmap\ and \Archeops\ data. In addition
to the above, many other related models and analyses can be found in
the literature.\\ 

\indent We propose in this paper a method to study and constrain the
synchrotron and thermal dust polarized emissions using the \Planck\
satellite observations that are currently underway. Using realistic
simulations of the \Planck\ polarized data we forecast the expected
constraints on a 3D model of the Galactic magnetic field and the Galactic
matter distribution. This paper is structured as follows: in Section~\ref{3dgal_model} we describe the models we used for the
polarized components of the Galactic diffuse emissions and in Section 3 we present the simulations
of data
used in this analysis. In Section~\ref{gal_comp} we describe our method to
constrain these Galactic foreground emission models and we discuss
the results in Section~\ref{discuss}. Conclusions are presented in Section~\ref{conc}.

\section{Models of polarized Galactic emissions}
\label{3dgal_model}

\indent A realistic model of synchrotron and thermal dust emissions
can be constructed from a 3D description of the Galaxy including the
matter distribution and the magnetic field structure. Following
\cite{fauvet}, we calculate the Stokes parameters
I, Q and U of the emerging polarized Galactic emissions along the
line-of-sight {\bf n}.\\
For the synchrotron emission, in galactocentric cylindrical coordinates $(r,\phi,z)$ we use the following model~\citep{fauvet}:

\begin{eqnarray*}
\label{eq:map_sync}
 I_{\mathrm{sync}}(\mathrm{{\bf n}}) &=& I_{\mathrm{Has/ff}}(\mathrm{{\bf n}}) \left(\frac{\nu_s}{0.408}\right)^{\beta_s},\\
Q_{\mathrm{sync}}(\mathrm{{\bf n}}) &=& I_{\mathrm{Has/ff}}(\mathrm{{\bf n}})  
\left(\frac{\nu_s}{0.408}\right)^{\beta_s}  \nonumber \\ && . \frac{\int \cos(2\gamma({\bf n},s))p_s
  \left(B_l^2({\bf n},s) + B_t^2({\bf n},s) \right)\mathrm{n}_{\mathrm{CRE}}(r,z)ds}{\int \left(B_l^2({\bf n},z) + B_t^2({\bf n},s) \right)\mathrm{n}_{\mathrm{CRE}}(r,z)ds}, \\ 
U_{\mathrm{sync}}(\mathrm{{\bf n}}) &=& I_{\mathrm{Has/ff}}(\mathrm{{\bf n}})
\left(\frac{\nu_s}{0.408}\right)^{\beta_s} \nonumber \\ && . \frac{\int \sin(2\gamma({\bf n},s))p_s
  \left(B_l^2({\bf n},s) + B_t^2({\bf n},s) \right)\mathrm{n}_{\mathrm{CRE}}(r,z)ds}{\int \left(B_l^2({\bf n},s) + B_t^2({\bf n},s) \right)\mathrm{n}_{\mathrm{CRE}}(r,z)ds},
\end{eqnarray*}

\noindent where $B_n(\mathrm{{\bf n}},s)$ is the magnetic component along the
line-of-sight {\bf n}, and $B_l(\mathrm{{\bf n}},s)$ and $B_t(\mathrm{{\bf n}},s)$ the magnetic field components
on a plane perpendicular to the line-of-sight. The polarization fraction $p_s$ is set to
75\%~\citep{ribicki}. The polarization angle $\gamma(\mathrm{{\bf n}},s)$ is given by :

\be
\centering
\gamma(\mathrm{{\bf n}},s) = \frac{1}{2} \arctan
\left(\frac{2B_l(\mathrm{{\bf n}},s)\cdot B_t(\mathrm{{\bf n}},s)}{B^2_l(\mathrm{{\bf n}},s)- B^2_t(\mathrm{{\bf n}},s)}\right)
\ee

\noindent The distribution of relativistic electrons,
$\mathrm{n}_{\mathrm{CRE}}$, is described in details in section~\ref{dens_matter}. $I_{Has/ff}$ is a
template temperature map obtained from the 408 MHz all-sky continuum
survey~\citep{haslam}. This map is also included bremsstrahlung
(free-free) emission. To substract this component we used the
\Wmap\ K-band free-free foreground map generated from the maximum entropy method
(MEM)~\citep{hinshaw07, bennett2003a}. Note that this template is not
necessarily realistic; \cite{alves} have shown with radio
recombination lines that in at least one region in the Galactic plane,
this model appears to overestimate the amount of free-free. However,
that will not have any impact in the following analysis because we
used the same template in the simulated data and in the fitted models
and the uncertainty at 408 MHz is very small compared to the
synchrotron amplitude. The free-free map is then extrapolated to 408 MHz
assuming a power law dependance as in~\cite{dickinson}. The spectral index $\beta_s$ used to
extrapolate maps at various frequencies is a free parameter of the
model and will be discussed later.\\

 For the thermal dust emission we used the following model~\citep{fauvet}: \\
\begin{eqnarray*}
\centering
 I_{\mathrm{dust}}(\mathrm{{\bf n}}) &=& \mathrm{I}_{\mathrm{fds}}(\mathrm{{\bf n}}) \left(\frac{\nu_d}{353} \right)^{\beta_d},\\
Q_{\mathrm{dust}}(\mathrm{{\bf n}}) &=& \mathrm{I}_{\mathrm{fds}}(\mathrm{{\bf n}}) \left(
  \frac{\nu_d}{353}\right)^{\beta_d} \nonumber \\ && . \frac{\int \cos(2 \gamma(\mathrm{{\bf n}},s))
  \sin^2(\alpha) f_{\mathrm{norm}}p_d \mathrm{n}_{\mathrm{dust}}(r,z)ds}{\int \mathrm{n}_{\mathrm{dust}}(r,z)ds},\\
U_{\mathrm{dust}}(\mathrm{{\bf n}}) &=& \mathrm{I}_{\mathrm{fds}}(\mathrm{{\bf n}})
\left(\frac{\nu_d}{353}\right)^{\beta_d} \nonumber \\ && . \frac{\int \sin(2 \gamma(\mathrm{{\bf n}},s))
  \sin^2(\alpha) f_{\mathrm{norm}}p_d \mathrm{n}_{\mathrm{dust}}(r,z)ds}{ \int \mathrm{n}_{\mathrm{dust}}(r,z)ds} ,
\end{eqnarray*}

\noindent where the dust polarization fraction $p_d$ is set to 10
\%~\citep{ponthieu2005}, and $\mathrm{n}_{\mathrm{dust}}(r,z)$ is the dust
grain distribution discussed in section~\ref{dens_matter}. The $sin^2(\alpha)$ term accounts for
geometrical supression and $f_{norm}$ is an
empirical factor which accounts for the misalignment between dust
grains and the magnetic field lines (see~\cite{fauvet} for
details). The reference map, $\mathrm{I}_{\mathrm{fds}}$, is 
the \cite{finkbeiner} model 8 prediction based on the \Iras\
data~\citep{neugebauer} and on  the COBE/DIRBE data. The spectral index $\beta_d$ used to
extrapolate maps at various frequencies is a free parameter of the
model. This template seems to be a good representation of the
\Archeops\ data at 353 GHz as discussed in~\cite{macias}. Notice that
for the polarized \Planck\ frequencies, we are in the Rayleigh-Jeans
domain and therefore, a power-law approximation for the dust intensity
can be used.\\

\subsection{3D model of the Galaxy}
\label{3dgal_model}

\indent We describe here our 3D model of the Galaxy.

\subsubsection{Matter density}
\label{dens_matter}

\indent We consider an exponential distribution of
relativistic electrons $\mathrm{n}_{\mathrm{CRE}}$ on the Galactic
disk motivated by~\cite{drimmel}:

\be
\centering
\mathrm{n}_{\mathrm{CRE}}(r,z) =  n_{0,e} \cdot \frac{e^{-\frac{r}{\mathrm{n}_{\mathrm{CRE},r}}}}{\cosh^2(z/\mathrm{n}_{\mathrm{CRE},h})}
\ee

\noindent where $\mathrm{n}_{\mathrm{CRE,r}}$ is the scale radius of the distribution
and is a free
parameter of the model. The vertical scale height,
$\mathrm{n}_{\mathrm{CRE,h}}$, is set to 1 kpc. The value of $n_{0,e}$
is set to $6.4 \times 10^{-6}$cm$^{-3}$, following~\cite{sun}.\\
\indent The distribution of dust grains $\mathrm{n}_{\mathrm{dust}}$ is described 
in the same way that of relativistic electrons with:

\be
\centering \mathrm{n}_{\mathrm{dust}}(r,z) = n_{0,d} \cdot \frac{e^{-\frac{r}{\mathrm{n}_{d,r}}}}{\cosh^2(z/\mathrm{n}_{d,h})},
\ee

\noindent where the scale radius $n_{d,r}(r,z)$ is
also a free parameter of the model (see~\cite{fauvet}
for more details). The vertical scale height, $\mathrm{n}_{d,r}$, is set to 1 kpc. 

\subsubsection{Galactic magnetic field model}
\label{mg_field_model}

\indent The Galactic magnetic field model is composed of two
parts: a regular component and a turbulent component such that $\vec{B}
= \vec{B}_{reg}+ \vec{B}_{turb}$ (bold letters indicate vectorial quantities). We include only the isotropic 
part of the turbulent magnetic field and no anisotropic/ordered component (see \cite{jaffe}).  As in \cite{fauvet}, our regular component is then equivalent to the sum of what Jaffe et al. (2010) call the coherent and ordered fields. For this analysis of synchrotron and dust emission only, the distinction is irrelevant. For the regular
component we consider a Modified Logarithmic Spiral model (MLS), presented
in~\cite{fauvet} and based on the
\Wmap\ model~\citep{page2007}. In cylindrical coordinates $(r,\phi,z)$ it reads :\\

\begin{eqnarray}
\centering
\mathbf{\vec{B}_{reg}}(\mathbf{r})&=& B(\mathbf {r})[ \cos(\phi+\beta) \ln \left( \frac{r}{r_0} \right) \sin(p) \cos(\chi(r) ) \cdot \mathbf{u_r} \nonumber \\&&- \cos(\phi+\beta) \ln \left( \frac{r}{r_0} \right) \cos(p) \cos(\chi(r)) \cdot \mathbf{u_{\phi}}  \nonumber \\&&+ sin(\chi(r)) \cdot \mathbf{u_z}] ,
\end{eqnarray}

\noindent where the pitch angle, $p$, is a free
parameter of the model and $\beta=1/\tan(p)$.  The scale radius $r_0$ is
set to 7.1 kpc and $\chi(r) = \chi_0(r)(z/z_0)$ is the vertical
scale height, with $\chi_0 = 22.4$ degrees and $z_0 = 1$kpc. 
 The intensity of the regular field is fixed using pulsar Faraday rotation measurements by \cite{han2006}:
\be
B(r) = B_0 \ e^{-\frac{r-R_{\odot}}{R_B}}
\ee

\noindent where the large-scale field intensity at the Sun position is $B_0=2.1 \pm 0.3 \mu
G$ and $R_B = 8.5 \pm 4.7 kpc$. The distance
between the Sun and the Galactic center, $R_{\odot}$ is set to 8 kpc
(\cite{eisenhauer, reid}).

\subsubsection{Turbulent component}

\indent In addition to the large-scale Galactic magnetic field,
Faraday rotation measurements on pulsars in our vicinity have revealed
a turbulent component \citep{lyne,han2004} with an amplitude
estimated to be of the same order of magnitude as that of the regular
one \citep{han2006}.\\
 
\indent The turbulent magnetic field is assumed to be a 3D anisotropic
gaussian random vectorial field and it is fully determined by a
spherically symmetric power spectrum in the Fourier domain. 
Indeed, the magnetic energy $E_B(k)$ associated with the turbulent
component can be described by a power spectrum of the form (\cite{han2004, han2006})
\be       
\centering
E_B(k) = C \left(\frac{k}{k_0}\right)^{\alpha}
\label{eq_pw_ko}
\ee

\noindent where $\alpha = -0.37$ and $C = (6.8 \pm 0.3) \cdot 10^{-13}\,\mathrm{erg\,cm^{-3}\,kpc}$.
More complex models of this anisotropic turbulent component have been
proposed by~\citep{higdon, cho2002} and~\cite{cho2010} but they are beyond the
scope of this paper. Note also that we do not consider here the
so-called
ordered turbulent component of the Galactic magnetic field as discussed
in~\cite{jaffe}.

\section{Simulated data}
\label{simu}

\indent We have performed simulations of the \Planck\ polarized
observations. We considered all the polarized channels for
the LFI~\citep{bersanelli, mandolesi, mennella} and
HFI~\citep{lamarre, cote} instruments at 30, 44, 70, 100, 143, 217 and 353
GHz for two full-sky surveys (14 months). We did not use the total
intensity maps in this analysis. For this 
set of simulations we generated full-sky maps in the HEALPix
pixelisation scheme~\citep{gorski} at $N_{side}=128$. The input
templates are simply degraded to that resolution using the HEALPix tools. \\
\indent At a given observation frequency, $\nu$, we consider that the
polarized observations ($X^{\nu}$) are the sum of the synchrotron
($X^{\nu}_{\mathrm{sync}}$) and the thermal dust emissions ($X^{\nu}_{\mathrm{dust}}$), where $X$ is the Q or U Stokes parameter. We also
add a CMB contribution ($X^{\nu}_{CMB}$) and noise ($X^{\nu}_N$) so
that can finally write: $X^{\nu}= X^{\nu}_{\mathrm{dust}} + X^{\nu}_{\mathrm{sync}} + X^{\nu}_N +X^{\nu}_{\mathrm{CMB}}$

\subsection{Polarized Galactic emission components}
\label{polar_simu}

\indent The polarized Galactic emissions are simulated using the 3D
model described in Section~\ref{3dgal_model}. The parameters
of the 3D model of the Galaxy have been taken from~\cite{fauvet}
by comparison with the available \Archeops~\citep{benoit2004a} and \Wmap\ 5-years
data~\citep{hinshaw}. The pitch
angle is set to -30 degrees and the radial widths of the distributions of dust
grains $n_{d,r}$ and
ultrarelativist electrons $n_{CRE,r}$ are set to 3 kpc.
We computed simulated data using a model of Galactic magnetic field
without (\emph{Simu I}) or with a turbulent component (\emph{Simu
  II}). In the second case we set the amplitude of the turbulent
component to $A_{turb}=0.25 \times B(r)$. The simulations are extrapolated to each of the
\Planck\ frequencies assuming constant spectral indices in
  frequency and using two different configurations, either spatially constant (\emph{Simu Cst}) or
variable (\emph{Simu Var}) accross the sky.  We have
  produced at least 10 simulations for each of the different configurations.\\

\indent In the \emph{Simu Cst} case we set the value of
the synchrotron spectral index to $\beta_s = -3.0$.  Concerning
  the thermal dust emission we test both upper and lower expected
  limits of the spectral index. We tested 1.4 as a lower limit, obtained
  by comparison of a grey body law model to the \Archeops\ and \Iris\
data in the galactic plane. The upper limit tested is 2, following
results obtained at high latitude by~\cite{boulanger} using the \Firas\
and \Dirbe\ data.\\

\begin{figure*}
\centering
\includegraphics[height=9cm,width=6cm, angle
=90]{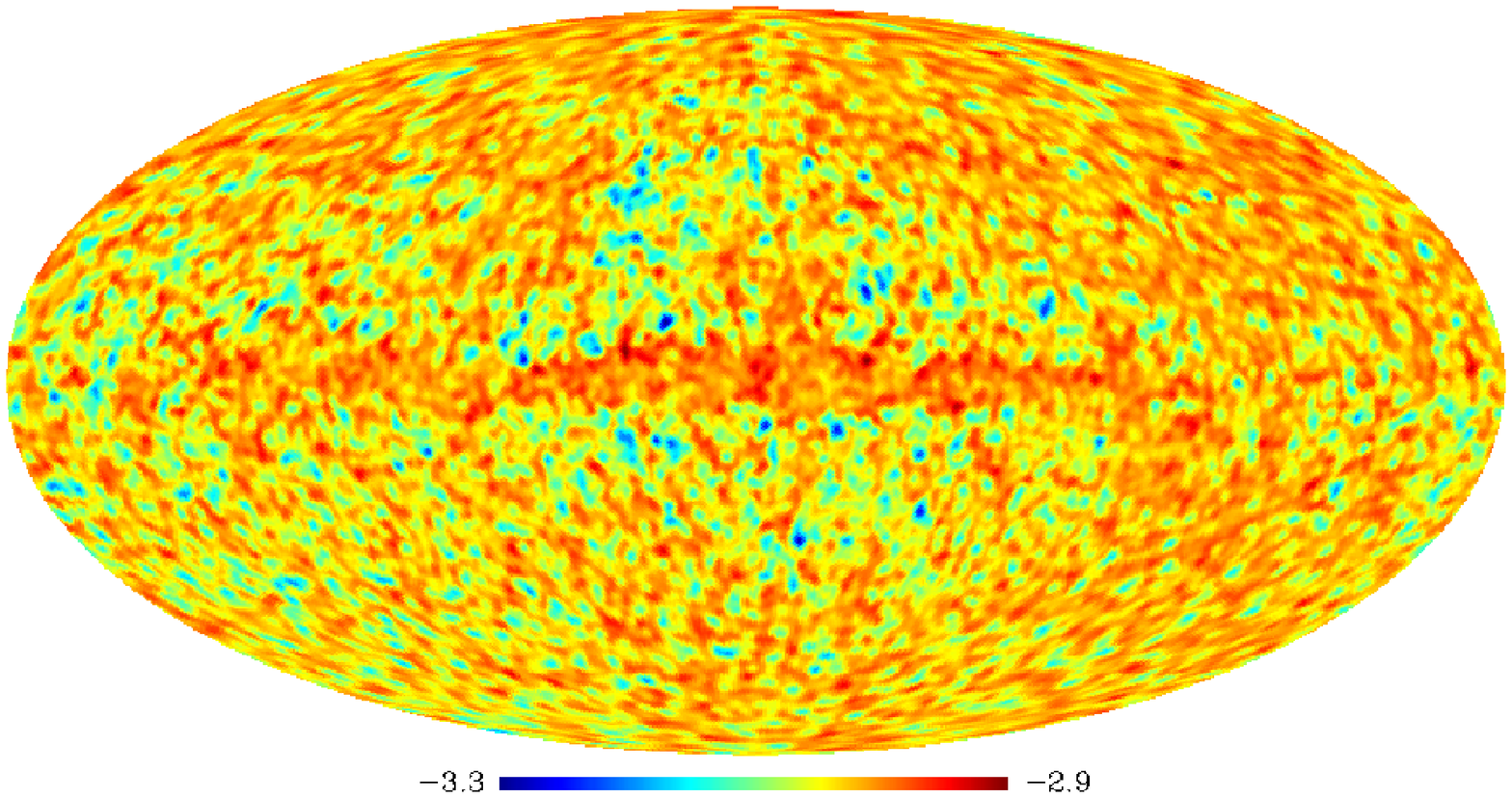}\includegraphics[height=9cm,width=6cm, angle
=90]{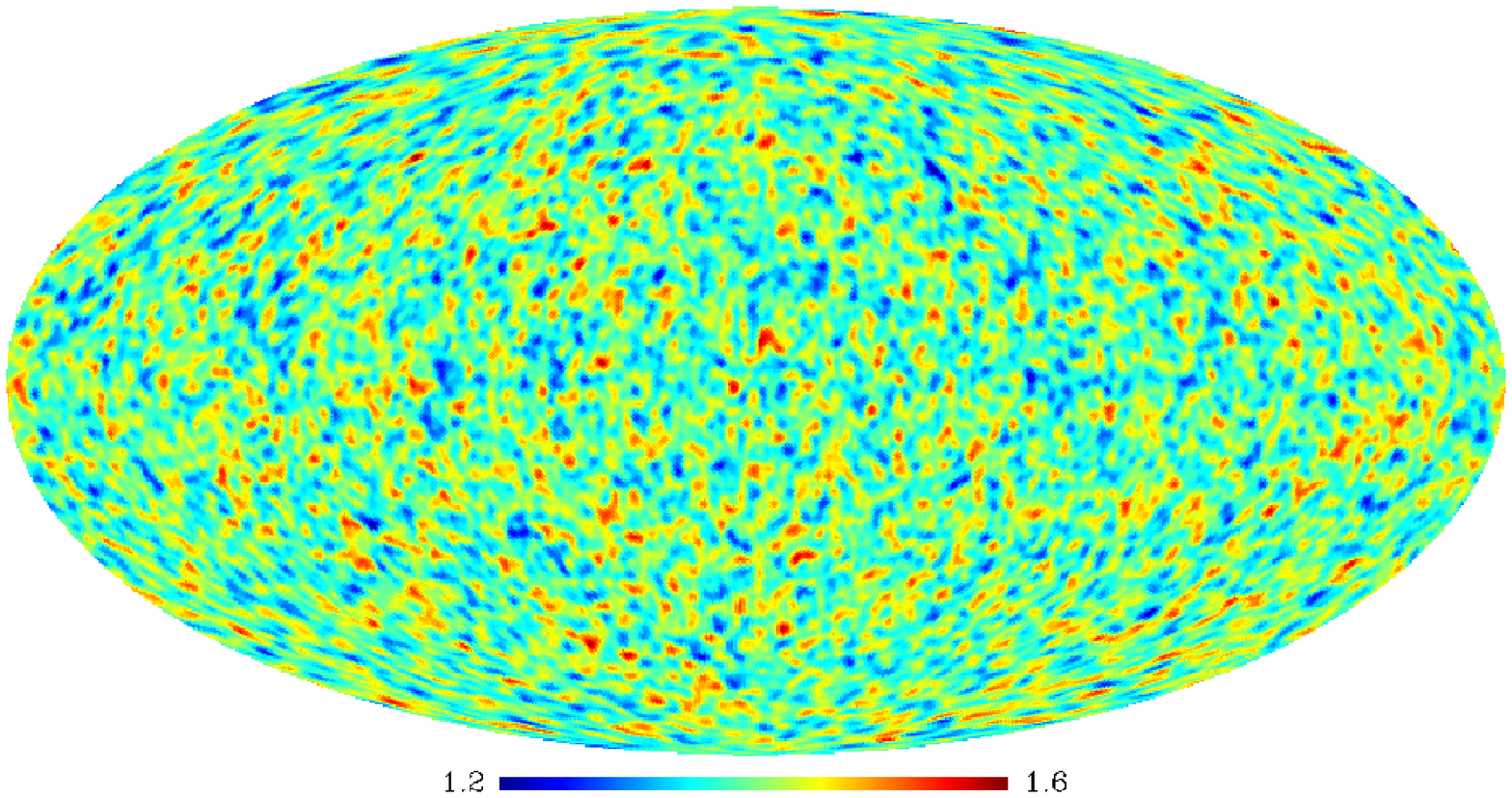}\caption{Spatial variations of the spectral indices for the simulated synchrotron (\emph{left}) and thermal dust Galactic polarized emissions (\emph{right}).}\label{beta_var}
\end{figure*}

\indent For the \emph{Simu Var} case,
the maps of $\beta_s$ and $\beta_d$ are built in the following way. For the synchrotron emission we used the map generated using
a
MCMC fit by~\cite{kogut} at 23 GHz applying some corrections. We cut
all pixels where $\beta_s < -3.5$ and $\beta_s > -2.8$, following 
the results in~\cite{page2007, sun, fauvet}, and assumed for those
pixels $\beta_s = -3.0 + 0.1*n$, where $n$ is a gaussian random
normal distribution.  Note that $\beta_s$ is the typical average value found in
  the literature. It is also important to clarify that this spectral index map was
  not derived from the MEM analysis of the \Wmap\ data. However, it
  can be used as a fair representation of the variability of the
  spectral index. The resulting synchrotron spectral index map is shown in the left
panel of Figure~\ref{beta_var}. \\
\indent To simulate the spatial variations of the spectral index of
the thermal dust
emission we assumed a random gaussian distribution with a mean value
of $\beta_{d,mean}=1.4$ or $\beta_{d,mean}=2.0$ and a standard deviation of 0.3. The right panel of Figure~\ref{beta_var}
shows the resulting thermal dust spectral index map. Table~\ref{kindsimu}
summarizes the resulting types of simulations that we carried out.

\begin{table*}
\begin{center}
\caption{Summary of the different types of simulations considered in this paper.\label{kindsimu}}
\vspace{0.3cm}

\begin{tabular}{|c|c|c|} \hline
Simulation       & $\beta_{s,d}$ & $A_{turb}$ \\\hline
\multirow{2}{*}{Simu 1}  & constant  & $0.0 \times B_{reg}$        \\\cline{2-3}
                         & variable  & $0.0 \times B_{reg}$         \\\cline{2-3}
\hline                      
\multirow{2}{*}{Simu 2}  &  constant  & $0.25 \times B_{reg}$        \\\cline{2-3}
                          & variable   & $0.25 \times B_{reg}$       \\\cline{2-3}
\hline
\end{tabular}
\end{center}
\end{table*}

\subsection{CMB emission}

\indent We produced maps at $N_{side}=128$ of the expected CMB signal
at all the \Planck\ frequencies. We used
CAMB~\citep{lewis} to compute the CMB temperature and polarization
angular power spectra for the \Wmap\ $\Lambda CDM$ best fit model as estimated by~\citep{komatsu}. We
also took into account gravitational lensing effects and assumed a 
tensor-scalar ratio, $r$, of 0.1~\citep{efstathiou}.

\subsection{Noise}

\indent \Planck\ noise maps for each of the frequency channels were
computed using the mean sensitivity per pixel given in
Table~\ref{sigma_planck}. We assumed isotropic random gaussian noise
accross the sky. This is not actually the case for the real
  \Planck\ scanning strategy~\citep{dupac}.  However this should
not have any impact on the presented results as we are mostly signal dominated.

\begin{table*}
\begin{center}
\caption{Average 1$\sigma$ sensitivity per pixel (a square whose side
  is the FWHM extent of the beam) in thermodynamic
  temperature units, achievable after 2 full sky surveys (14 months) by \Planck.~\citep{bluebook}}\label{sigma_planck}
\vspace{0.3cm}
\begin{tabular}{|c|c|c|c|c|c|c|c|} \hline
  Center frequency [GHz] & 30 & 44 & 70 & 100 & 143 & 217 & 353 \\\hline
($\Delta T/T$) polarization [$\mu K/K$] & 2.8 & 3.9 & 6.7 & 4.0 & 4.2 & 9.8 & 29.8 \\\hline
Angular resolution [arcmin FWHM] & 33 & 24 & 14 & 10 & 7 & 5 & 5 \\\hline
\end{tabular}
\end{center}
\end{table*}

\section{Method}
\label{gal_comp}

\begin{table*}
\begin{center}
\caption{Parameters of the 3D Galactic polarized diffuse emission models.}\label{param_tab}
\vspace{0.3cm}

\begin{tabular}{|c|c|c|} \hline
Parameters      &   Range  &   Binning  \\\hline
 $p$ (deg)      &  $[-80.0,80.0]$     & $10.0$   \\\hline
 $A_{turb}$     &   $[0,2.5]*B_{reg}$ &   $0.125$ \\\hline 
 $n_{CRE,r}$ (kpc)    & $[1.0,20.0]$        &   $1$ \\\hline
 $n_{d,r}$ (kpc)    & $[1.0,20.0]$        &   $1$ \\\hline
 $\beta_s$      & $[-4.3,-2.4]$       & $0.1$ \\\hline
 $\beta_d$      & $[2.0,4.0]$         & $0.1$ \\\hline

\end{tabular}
\end{center}
\end{table*}

\indent Following~\cite{fauvet}, in order to compare the models of Galactic polarized emissions to the
\Planck\ data simulations we computed Galactic profiles in polarization
using the set of latitude bands (in degrees) $[0,30]$, $[30,60]$, $[60,90]$, $[90,120]$,$[120,180]$, $[180,270]$, $[270,330]$, $[330,360]$. Note that the intensity profiles are not used in
this paper as the intensity maps were constructed using fixed templates.\\

\indent Galactic latitude profiles for the diffuse Galactic polarized
emission models were computed for a grid of models obtained by varying the pitch
angle, $p$, the turbulent component amplitude, $A_{turb}$, the radial
scale for the distribution of electrons, $n_{CRE,r}$ and of dust grains
$n_{d,r}$, and the spectral indices $\beta_s$ and $\beta_d$. The latter
were assumed to be spatially constant accross the sky. All the others parameters were set to the values
proposed in section~\ref{polar_simu}.\\

\indent We compared the simulated data sets to the Galactic emission
models using a likelihood analysis where the log-likelihood function is
given by: \\ 

\be
- log \mathcal{L} = \nonumber \\ 
\frac{\sum_{\nu=0}^{nfreq-1}\sum_{i=0}^{nlon-1}\sum_{n=0}^{nlat-1}(S^{\nu}_{i,X}(n)-M^{\nu}_{i,X}(n))^2}{\sigma^{\nu}_{i,X}(n)^2+\sigma^{\nu,turb}_{i,X}(n)^2}
\ee

\noindent where the $X$ are the Stokes parameters $Q$ and $U$, and $i$ and $n$
index the longitude bands and the latitude bins, respectively.
$S^{\nu}_{i,X}(n)$ and $M^{\nu}_{i,X}(n)$ are the set of
simulations and models respectively for the polarization state $X$, $i$ longitude
band and $n$ longitude bin. The term $\sigma^{\nu}_{i,X}(n)^2$ is the error
associated to $S^{\nu}_{i,X}(n)$ computed from the standard
deviation of the data samples in each of the latitude bins. Note
that it accounts both for the noise and signal dispersion within the bin. The term $\sigma^{\nu,turb}_{i,X}(n)$
 accounts for the additional variance due to the turbulent component of the
 magnetic field.  Indeed, as the magnetic field is considered to be a
 random distribution, we need to take into account in the likelihood
 function an extra correlation matrix. We approximated this matrix to a
 diagonal one. We used 10 simulations of the Galactic turbulent
 contribution at each \Planck\ frequency band to estimate $\sigma^{\nu,turb}_{i,X}(n)$. 
Note that the latter term is proportional to $A_{turb}$ and also to the extrapolation term, $(\frac{\nu}{\nu_{ref}})^{\beta}$, both for the synchrotron and thermal dust components.
This may introduce a small bias in the A$_{turb}$, $\beta_{s}$ and $\beta_{d}$ parameters. Increased turbulence can be balanced by a steeper spectral
index and therefore it is possible to lower the resulting $\chi^2$ for some parameter combinations.  We will see, however, that the impact of this bias on the results remains small.

\begin{table*}
\begin{center}
\caption{Best-fit parameters for the Galactic polarized emission
  models in the case of $\beta_d=1.4$ in the simulated data.   In the case of simulations with turbulence, \emph{Simu II}, we give for each
    parameter the variance among the set of simulation results.}\label{param}
\begin{tabular}{|c|c|c|c|c|} \hline
Simulation & \multicolumn{2}{c|}{Simu I} & \multicolumn{2}{c|}{Simu II} \\\hline\hline
$\beta_{simu}$ & Cst & Var & Cst & Var \\\hline
$A_{turb}$ & $<0.1$ & $<0.1$ &  $0.25^{+0.2}_{-0.1}$ &  $0.25^{+0.2}_{-0.1}$ \\\hline
$ p(deg)$ & $-30^{+4}_{-6}$ & $-30^{+4}_{-6}$ & $-30^{+8}_{-14}$ (5) & $-30^{+8}_{-14}$ (5) \\\hline
$n_{CRE,r}$ & $3^{+1.5}_{-1}$ & $3^{+1.5}_{-1}$ & $12^{+6}_{-8}$ (4.5) & $12^{+6}_{-8}$ (4.5) \\\hline
$n_{d,r}$ &  $3^{+10}_{-1}$ & $3^{+10}_{-1}$ & $ < 16$ (5.1) & $ 12{+6}_{-8} $  (4.5) \\\hline
$\beta_s$ & $-3.0^{+0.05}_{-0.1}$ & $-3.0^{+0.05}_{-0.1}$ & $-3.1^{+0.1}_{-0.7}$  ( $<$ 0.05)  & $-3.1^{+0.1}_{-0.7}$  ( $<$ 0.05)  \\\hline
$\beta_d$ &  $1.4 \pm 0.1$  & $1.4^{+0.2}_{-0.4}$ & $1.5^{+0.5}_{-0.3}$ (0.05) &
$1.5^{+0.5}_{-0.3}$ (0.05) \\\hline
\end{tabular}
\end{center}
\end{table*}

\begin{table*}
\begin{center}
\caption{Best-fit parameters for the Galactic polarized emission
  models in the case of $\beta_d=2.0$ in the simulated data. In the case of simulations with turbulence, \emph{Simu II}, we give for each
    parameter the variance among the set of simulation results.}\label{param_2}
\begin{tabular}{|c|c|c|c|c|} \hline
Simulation & \multicolumn{2}{c|}{Simu I} & \multicolumn{2}{c|}{Simu II} \\\hline\hline
$\beta_{simu}$ & Cst & Var & Cst & Var \\\hline
$A_{turb}$ & $<0.1$ & $<0.1$ &  $0.35^{+0.2}_{-0.1}$ (0.06) &  $0.35^{+0.2}_{-0.1}$(0.06) \\\hline
$ p(deg)$ & $-30^{+4}_{-6}$ & $-30^{+4}_{-6}$ & $-30^{+12}_{-14}$ (7.5) & $-30^{+12}_{-14}$ (7.5) \\\hline
$n_{CRE,r}$ & $3^{+1.5}_{-1}$ & $3^{+1.5}_{-1}$ & $12^{+6}_{-8}$ (4.1) & $12^{+6}_{-8}$ (4.1) \\\hline
$n_{d,r}$ &  $3^{+10}_{-1}$ & $3^{+10}_{-1}$ & $ < 16$ (2) & $ 12^{+6}_{-8} $  (2) \\\hline
$\beta_s$ & $-3.0^{+0.05}_{-0.1}$ & $-3.0^{+0.05}_{-0.1}$ & $-3.1^{+0.1}_{-0.7}$  ( 0.05)  & $-3.1^{+0.1}_{-0.7}$  (0.05)  \\\hline
$\beta_d$ &  $2.0 \pm 0.1$  & $2.0^{+0.2}_{-0.4}$ & $2.2^{+0.5}_{-0.3}$ (0.05) &
$2.3^{+0.5}_{-0.3}$ (0.05) \\\hline
\end{tabular}
\end{center}
\end{table*}

\section{Results and discussion}
\label{discuss}

\indent The expected constraints on the parameters of the polarized
Galactic emission models using the simulated \Planck\ data are given in
Tables~\ref{param} and \ref{param_2} for the various sets of simulations considered.

\subsection{Simu I}

\begin{figure*}
\centering
\includegraphics[height=15cm,width=15cm]{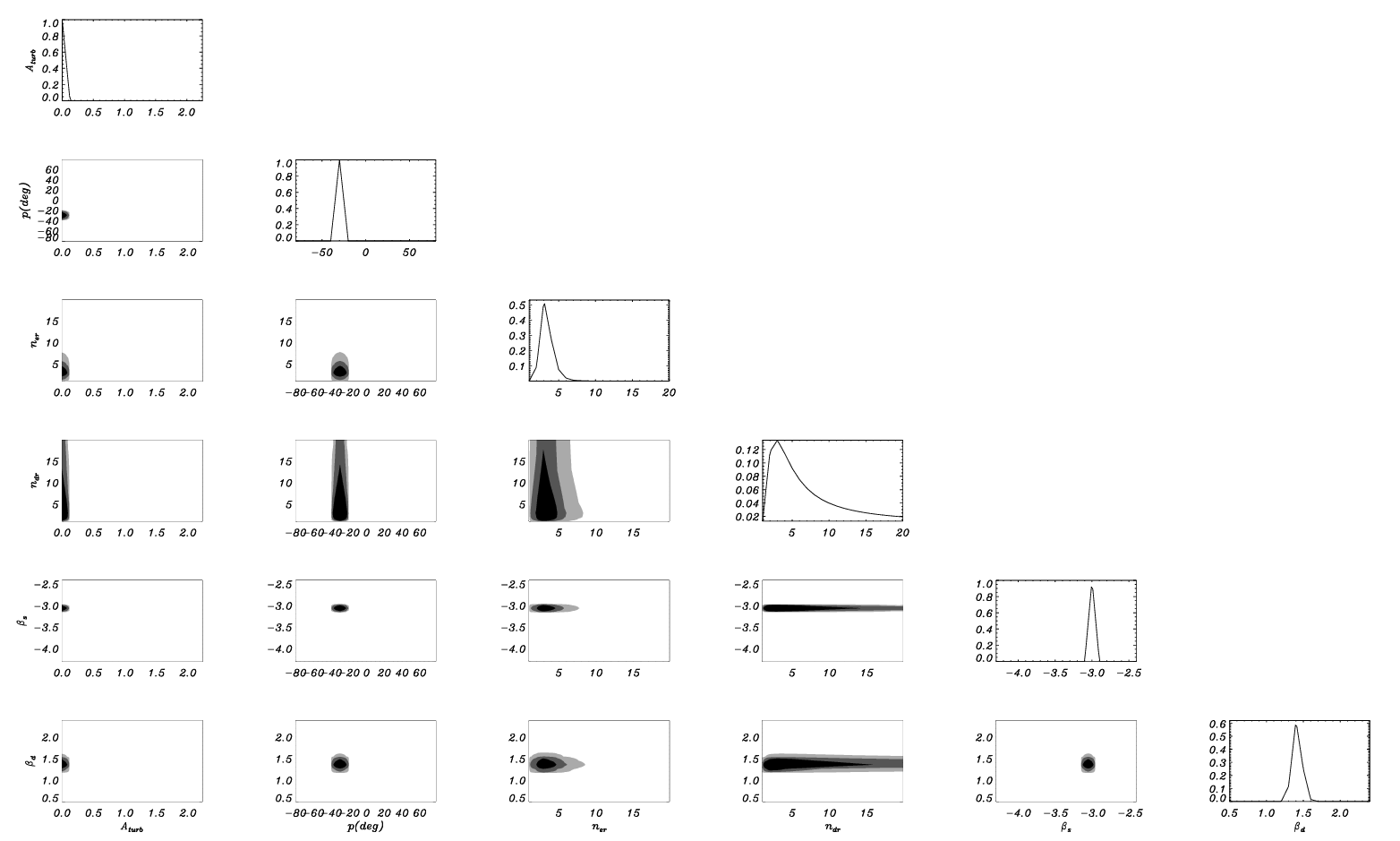}\caption{Marginalised
  likelihood in 1 and 2D for the parameters $A_{turb}$ and $p$,
  $n_{CRE,r}$ , $n_{d,r}$, $\beta_s$ and $\beta_d$ for the (\emph{Simu
    I Cst}) case and we present the 1 (68.8\%), 2 (95.4\%) and
3$\sigma$ (98\%) confidence level contours. The values of the
parameters included in the simulated data are set to $A_{turb}$=0,
$p = -30^{\circ}$, $n_{CRE,r} = n_{d,r} = 3$kpc, $\beta_s = -3.0$,
$\beta_d = 1.4$ respectively.\label{like_cst}}
\end{figure*}

\indent The constraints obtained for the \emph{Simu I} cases
are presented in the first line of Tables~\ref{param} and~\ref{param_2}. The associated
marginalized likelihood in 1 and 2D are shown in
Figure~\ref{like_cst} for the parameters $A_{turb}$, $p$, $n_{CRE,r}$,
$n_{d,r}$, $\beta_s$ and $\beta_d$, and we present the 1, 2 and
3$\sigma$ confidence level contours. We can see that there is no
correlation between the parameters. We are able to
tightly constrain all the parameters of the Galactic emission
models. Furthermore, we can see that the best fit values coincide with the parameters used in
the input simulations. Therefore, there is no indication of bias in
the method. We can see that spatial variations of the
spectral indices do not disturb the constraints on the
parameters. Indeed the expected constraints on all the parameters of the models,
including the values of spectral indices, are unchanged. This could be
explained by the fact that we use a Galactic profile based comparison
 which is not very sensitive to pixel-to-pixel variations but rather to
 global features. The constraint on the dust grain density
parameter is weaker than that for the relativistic electrons density. This is
probably due to the fact that the synchrotron is dominant at 
the \Planck\ frequencies.

\subsection{Simu II}

\begin{figure*}
\centering
\includegraphics[height=15cm,width=15cm]{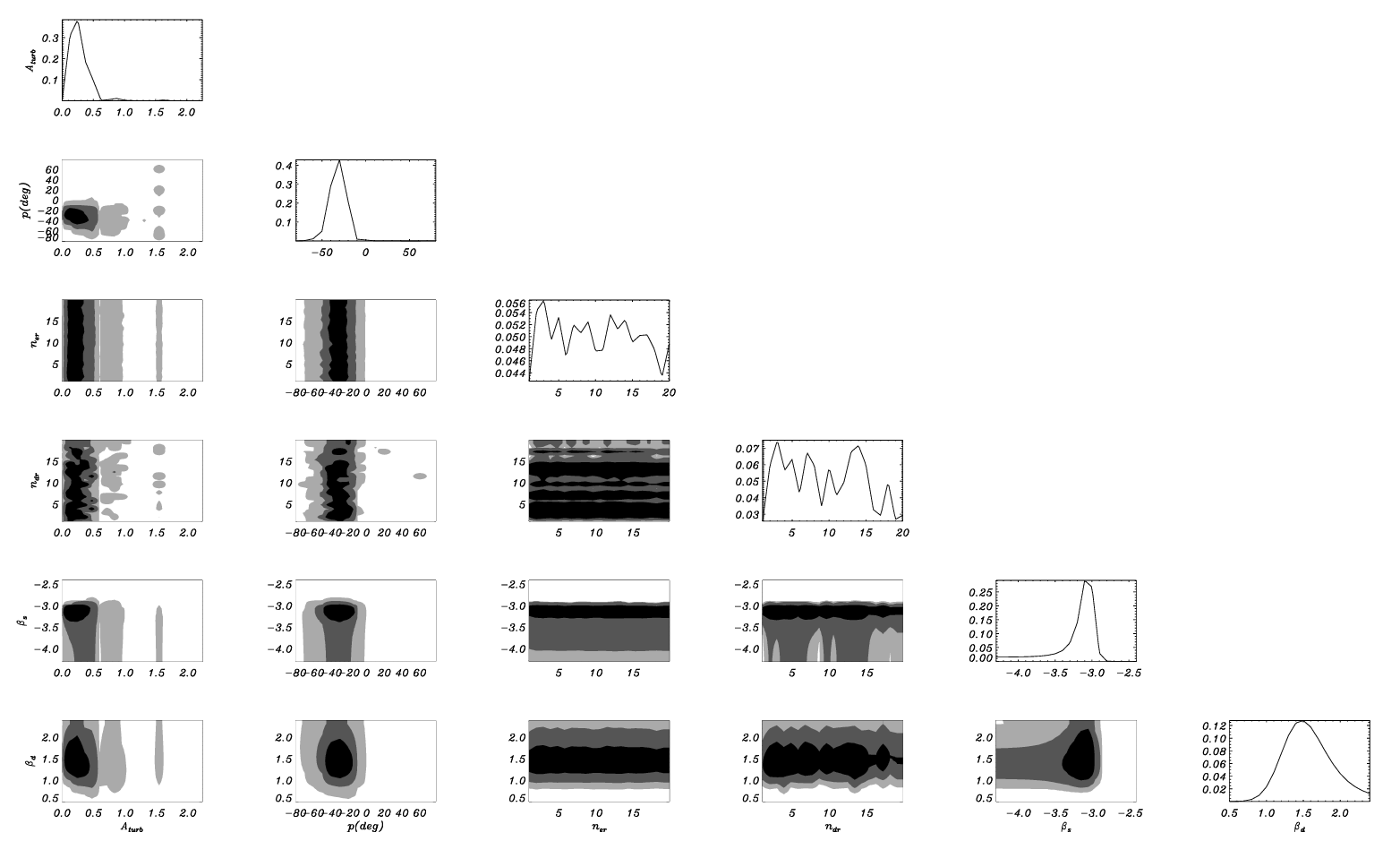}\caption{Marginalised  likelihood in 1 and 2D for the parameters $A_{turb}$ and $p$,
  $n_{CRE,r}$ , $n_{d,r}$, $\beta_s$ and $\beta_d$ for the (\emph{Simu
    II Var}) case and we present the 1 (68.8\%), 2  (95.4\%) and
3$\sigma$ (98\%) confidence level contours. The values of the
parameters included in the simulated data are set to $A_{turb}$=0,
$p = -30^{\circ}$, $n_{CRE,r} = n_{d,r} = 3$kpc, $\beta_s = -3.0$,
$\beta_d = 1.4$ respectively. \label{like_var}}
\end{figure*}

\indent The results concerning the \emph{Simu II} cases, i.e. those including a turbulent component, are 
summarized in the lines 3 and 4 of Tables~\ref{param} and~\ref{param_2}. The marginalized
likelihoods in 1 and 2D for the parameters $A_{turb}$, $p$,
$n_{CRE,r}$, $n_{d,r}$, $\beta_s$ and $\beta_d$ are shown in
figure~\ref{like_var} on which we present the 1, 2 and 3$\sigma$
confidence level contours. We observe that there is no correlation
between parameters, however, the constraints on the 
different parameters are weaker by a factor of 2 or more compared to the
\emph{Simu I} case. We observe a small bias on the best-fit values of
the two spectral indices, but it is within the 1$\sigma$ error bars;  it is related to the additional noise-like term added to the likelihood calculation by the turbulent component of the magnetic field. Furthermore, the
radial scales $n_{CRE,r}$ and $n_{d,r}$ are not constrained. A
 degeneracy between the matter distribution and the other parameters
 of the models is
 induced by the method of construction but is not visible in
the Figure~\ref{like_var}.  Nevertheless, an upper limit can be
set. Because the millimeter and
submillimeter data are not very sensitive to changes in those
parameters, as was discussed in~\cite{fauvet}, only an upper limit can be determined.

\section{Summary and Conclusions}
\label{conc}

\indent We proposed a method to estimate the expected constraints on
the Galactic diffuse polarized emissions and the Galactic magnetic
field at large scales using the \Planck\ data. With this aim, we
computed realistic simulations of the \Planck\ data at the polarized
frequency bands for two all-sky surveys. These simulations include
CMB, synchrotron and thermal dust emissions and instrumental
noise. For the synchrotron and thermal dust Galactic emissions we used
a coherent 3D model of the Galaxy describing the magnetic field
direction and intensity and the distribution of matter. The
relativistic electron and dust grain densities were modeled using
exponential distributions in galactocentric coordinates. For the
Galactic magnetic field we considered the Modified Logarithmic Spiral
model discussed in~\cite{fauvet}.\\

\indent We performed a likelihood analysis to compare the simulated
\Planck\ data to a set of models obtained by varying the pitch angle
of the regular magnetic field spatial distribution, the relative
amplitude of the turbulent magnetic field, the radial scale of the
electron and dust grain distributions as well as the extrapolation
indices of the synchrotron and thermal dust emissions. We are able to
set accurate constraints on most of the parameters considered. We have
also found that the observed spatial variations of the synchrotron and
thermal dust spectral indices should not affect our ability to recover
the other parameters of the model. The presence of a turbulent
component of the Galactic magnetic field decreases the discriminatory
power of the method for all parameters but only in the case of the
radial scales of the relativistic electron and dust grain
distributions  does it prevent an useful measurement. The small degree
of bias the results in simulations including the turbulent component
should not strongly affect the results with the real \Planck\ data in
polarization, since it remains small compared to the uncertainties.\\

\indent We can conclude that using the \Planck\
data we should be able to constrain simultaneously the parameters of
the models of synchrotron and thermal dust emissions. In particular, we expect to constraint the direction of the
Galactic magnetic field at large scales and the relative contributions
of the regular and the
istotropic turbulent component of the Galactic magnetic field without using 
external datasets. With respect to the current analysis, the
constraints on the dust grain density parameters could be improved
using also the total intensity data at the \Planck\ HFI channels, from 100
to 857 GHz. More generally, a more precise reconstruction of the matter
distribution in the Galaxy would require
on the one hand an improved modelling of the
ISM and on the other hand extra data sets like rotation measurements
of pulsars~\cite{han2004, sun}. These rotation measurement data along with total intensity 
should also help to constrain the ordered
turbulent Galactic magnetic field (see Jaffe et al, 2010), which has not been considered in
the current work.

\bibliographystyle{aa} 
\bibliography{biblio} 

\begin{thebibliography}{60}
\expandafter\ifx\csname natexlab\endcsname\relax\def\natexlab#1{#1}\fi

\bibitem[{Alves {et~al.}(2010)Alves, Davies, Dickinson, Davis, Auld,
  Calabretta, \& Staveley-Smith}]{alves}
Alves, M., Davies, R., Dickinson, C., {et~al.} 2010, MNRAS

\bibitem[{{Battistelli} {et~al.}(2006){Battistelli}, {Rebolo},
  {Rubin{\~o}-Martin}, {}Hildebrandt, {Watson}, {Guti{\`e}rrez}, \&
  {Hoyland}}]{battistelli2006}
{Battistelli}, E., {Rebolo}, R., {Rubin{\~o}-Martin}, J., {et~al.} 2006, \apj,
  645, 141

\bibitem[{Bennett {et~al.}(2003)Bennett, Halpern, Hinshaw, Jarosik, Kogut,
  Limon, Meyer, Page, \& Spergel}]{bennett2003a}
Bennett, C., Halpern, M., Hinshaw, G., {et~al.} 2003, \apjs, 148, 1

\bibitem[{Beno\^it {et~al.}(2004)Beno\^it, Ade, Amblard, Ansari, Aubourg,
  Bargot, Bartlett, Bernard, Bhatia, \& Blanchard}]{benoit2004a}
Beno\^it, A., Ade, P., Amblard, A., {et~al.} 2004, \aap, 424, 571

\bibitem[{Bersanelli {et~al.}(2010)Bersanelli, Mandolesi, N.~Butler, Villa,
  Aja, Artal, Artina, Baccigalupi, Balasini, Baldan, Banday, Bastia, \&
  Battaglia}]{bersanelli}
Bersanelli, Mandolesi, M., N.~Butler, R.C.~Mennella, A., {et~al.} 2010, A\&A,
  520

\bibitem[{Boulanger {et~al.}(1996)Boulanger, Abergel, Bernard, Burton,
  D{\'e}sert, Hartmann, Lagache, \& Puget}]{boulanger}
Boulanger, F., Abergel, A., Bernard, J.-P., {et~al.} 1996, \aap, 312, 181

\bibitem[{Brouw \& Spoelstra(1976)}]{brouw}
Brouw, N. \& Spoelstra, T. 1976, \aap Sup. S., 26, 129

\bibitem[{Cho \& Lazarian(2010)}]{cho2010}
Cho, J. \& Lazarian, A. 2010, \apj, 720, 1181

\bibitem[{Cho {et~al.}(2002)Cho, Lazarian, \& Vishniac}]{cho2002}
Cho, J., Lazarian, A., \& Vishniac, E.-T. 2002, \apj, 285, 109

\bibitem[{Cordes \& Lazio(2002)}]{cordes}
Cordes, J. \& Lazio, T. 2002, astro-ph/0207156

\bibitem[{Davis \& Greenstein(1951)}]{davis}
Davis, B. \& Greenstein, J. 1951, \apj, 114, 206

\bibitem[{D{\'e}sert {et~al.}(1990)D{\'e}sert, Boulanger, \&
  Puget}]{desert1998}
D{\'e}sert, F.-X., Boulanger, F., \& Puget, J.-L. 1990, \aap, 237, 215

\bibitem[{Dickinson {et~al.}(2003)Dickinson, Davies, \& Davis}]{dickinson}
Dickinson, C., Davies, R., \& Davis, R. 2003, MNRAS, 341, 369

\bibitem[{Drimmel \& Spergel(2001)}]{drimmel}
Drimmel, R. \& Spergel, D. 2001, \apj, 556, 181

\bibitem[{Duncan {et~al.}(1999)Duncan, Reich, Reich, \& Furst}]{duncan1999}
Duncan, A., Reich, P., Reich, W., \& Furst, E. 1999, \aap, 350, 447

\bibitem[{Dupac(2005)}]{dupac}
Dupac, X. \&~Tauber, J. 2005, \aap, 430, 363

\bibitem[{Efstathiou {et~al.}(2009)Efstathiou, Gratton, \& Paci}]{efstathiou}
Efstathiou, G., Gratton, S., \& Paci, F. 2009, MNRAS, 397, 1355

\bibitem[{Eisenhauer {et~al.}(2003)Eisenhauer, Schodel, Genzel, Ott, Tecza,
  Abuter, \& Eckart}]{eisenhauer}
Eisenhauer, F., Schodel, R., Genzel, R., {et~al.} 2003, \apj, 597, L121

\bibitem[{Fauvet {et~al.}(2010)Fauvet, Mac\'ias-P{\'e}rez, Aumont, D{\'e}sert,
  Jaffe, Banday, Tristram, Waelkens, \& Santos}]{fauvet}
Fauvet, L., Mac\'ias-P{\'e}rez, J.~F., Aumont, J., {et~al.} 2010, A\&A, 526,
  145

\bibitem[{Finkbeiner {et~al.}(1999)Finkbeiner, Davis, \& Schlegel}]{finkbeiner}
Finkbeiner, D.~P., Davis, M., \& Schlegel, D., J. 1999, \apj, 524, 867

\bibitem[{Gold {et~al.}(2009)Gold, Odegard, Weiland, Hill, Kogut, Bennett,
  Hinshaw, Chen, Dunkley, Halpern, Jarosik, Komatsu, Larson, Limon, Meyer,
  Nolta, Page, Smith, Spergel, Tucker, Wollack, \& Wright}]{gold}
Gold, B., Odegard, N., Weiland, J.~L., {et~al.} 2009, \apjs, 180, 265

\bibitem[{G{\'o}rski {et~al.}(2005)G{\'o}rski, Hivon, Banday, Wandelt, Hansen,
  Reinecke, \& Bartelmann}]{gorski}
G{\'o}rski, K., Hivon, E., Banday, A., {et~al.} 2005, \apj, 622, 759

\bibitem[{Han {et~al.}(2004)Han, Ferri{\`e}re, \& Manchester}]{han2004}
Han, J.~L., Ferri{\`e}re, K., \& Manchester, R.~N. 2004, \aap, 610, 820

\bibitem[{Han {et~al.}(2006)Han, Manchester, Lyne, Qiao, \& van
  Straten}]{han2006}
Han, J.~L., Manchester, R., Lyne, A., Qiao, G.~J., \& van Straten, W. 2006,
  \aap, 642, 868

\bibitem[{Haslam {et~al.}(1982)Haslam, Salter, Stoffel, \& Wilson}]{haslam}
Haslam, C., Salter, C., Stoffel, H., \& Wilson, W.~E. 1982, \aaps, 47, 1

\bibitem[{Higdon(1984)}]{higdon}
Higdon, J.-C. 1984, \apj, 285, 109

\bibitem[{Hildebrand {et~al.}(1999)Hildebrand, Dotson, Dowell, Schleuning, \&
  Vaillancourt}]{hildebrand1999}
Hildebrand, R.~H., Dotson, J., Dowell, C., Schleuning, D.~A., \& Vaillancourt,
  J.~E. 1999, \apj, 516, 834

\bibitem[{Hinshaw {et~al.}(2007)Hinshaw, Nolta, Bennett, Bean, Dor{\'e},
  Greason, Halpern, Hill, Jarosik, \& Kogut}]{hinshaw07}
Hinshaw, G., Nolta, M., Bennett, C., {et~al.} 2007, \apjs, 170, 288

\bibitem[{Hinshaw {et~al.}(2009)Hinshaw, Weiland, Hill, Odegard, Larson,
  Bennett, Dunkley, Gold, Greason, Jarosik, Komatsu, Nolta, Page, Spergel,
  Wollack, Halpern, Kogut, Limon, Meyer, Tucker, \& Wright}]{hinshaw}
Hinshaw, G., Weiland, J., Hill, R., {et~al.} 2009, \apjs, 180, 225

\bibitem[{Jaffe {et~al.}(2010)Jaffe, Leahy, Banday, Leach, Lowe, \&
  Wilkinson}]{jaffe}
Jaffe, T., Leahy, J., Banday, A., {et~al.} 2010, MNRAS, 401, 1013

\bibitem[{Jansson {et~al.}(2009)Jansson, Farrar, Waelkens, \&
  Ensslin}]{jansson}
Jansson, R., Farrar, G., Waelkens, A., \& Ensslin, T. 2009, JCAP, 7, 21

\bibitem[{Kogut {et~al.}(2007)Kogut, Dunkley, Bennett, Dor{\'e}, Gold, Halpern,
  Hinshaw, Jarosik, Komatsu, Nolta, Odegard, Page, Spergel, Tucker, Weiland,
  Wollack, \& Wright}]{kogut}
Kogut, A., Dunkley, J., Bennett, C., {et~al.} 2007, \apj, 665, 355

\bibitem[{Komatsu {et~al.}(2009)Komatsu, Dunkley, Nolta, Bennett, Gold,
  Hinshaw, Jarosik, Larson, Limon, Page, Spergel, Halpern, Hill, Kogut, Meyer,
  Tucker, Weiland, Wollack, \& L.}]{komatsu}
Komatsu, E., Dunkley, J., Nolta, M., {et~al.} 2009, \apjs, 180, 330

\bibitem[{Lamarre {et~al.}(2010)Lamarre, Puget, Bouchet, Ade, Benoit, Bernard,
  Bock, \& De~Bernardis}]{lamarre}
Lamarre, J.-M., Puget, J., Bouchet, F., {et~al.} 2010, A\&A, 520

\bibitem[{Lawson {et~al.}(1987)Lawson, Mayer, Osborne, \& Parkinson}]{lawson}
Lawson, K., Mayer, C., Osborne, J., \& Parkinson, M. 1987, MNRAS, 225, 307

\bibitem[{Lewis {et~al.}(2000)Lewis, Challinor, \& Lasenby}]{lewis}
Lewis, A., Challinor, A., \& Lasenby, A. 2000, \apj, 538, 473

\bibitem[{Lyne \& Smith(1989)}]{lyne}
Lyne, A. \& Smith, F. 1989, MNRAS, 237, 533

\bibitem[{Mac\'ias-P{\'e}rez {et~al.}(2007)Mac\'ias-P{\'e}rez, Lagache, Maffei,
  Ade, Amblard, Ansari, Aubourg, Aumont, Bargot, Bartlett, Benoit, Bernard,
  Bhatia, Blanchard, Bock, Boscaleri, Bouchet, Bourrachot, Camus, Cardoso,
  Couchot, de~Bernardis, Delabrouille, Desert, Dor{\'e}, Douspis, Dumoulin,
  Dupac, Filliatre, Fosalba, Ganga, Gannaway, Gautier, Giard, Giraud~Heraud,
  Gispert, Guglielmi, Hamilton, Hanany, Henrot~Versille, Hristov, Kaplan,
  Lamarre, Lange, Madet, Magneville, Marrone, Masi, Mayet, Murphy, Naraghi,
  Nati, Patanchon, Perdereau, Perrin, Plaszczynski, Piat, Ponthieu, Prunet,
  Puget, Renault, Rosset, \& Santos}]{macias}
Mac\'ias-P{\'e}rez, J., Lagache, G., Maffei, B., {et~al.} 2007, \aap, 467, 1313

\bibitem[{Mandolesi {et~al.}(2010)Mandolesi, Bersanelli, Butler, Artal,
  Baccigalupi, Balbi, Banday, Barreiro, Bartelmann, Bennett, Bhandari, A.,
  Borrill, Bremer, Burigana, \& Bowman}]{mandolesi}
Mandolesi, N., Bersanelli, M., Butler, R., {et~al.} 2010, A\&A, 520

\bibitem[{Menella {et~al.}(2011)Menella, Bersanelli, Butler, Cuttaia,
  D'Arcangelo, Davis, Frailis, Galeotta, Gregorio, Lawrence, Lowe, \&
  Mandolesi}]{mennella}
Menella, A., Bersanelli, M., Butler, R., {et~al.} 2011, A\&A submitted

\bibitem[{Neugebauer {et~al.}(1984)Neugebauer, Habing, van Duinen, Aumann,
  Baud, Beichman, Boggess, Clegg, \& de~Jong}]{neugebauer}
Neugebauer, G., Habing, H.~J., van Duinen, R., {et~al.} 1984, \apj, 278, 1

\bibitem[{Nolta(2009)}]{nolta2009}
Nolta, M. 2009, \apjs, 180, 296

\bibitem[{Page {et~al.}(2007)Page, Hinshaw, Komatsu, Nolta, Spergel, Bennett,
  Barnes, Bean, Dor{\'e}, Dunkley, Halpern, Hill, Jarosik, Kogut, Limon, Meyer,
  Odegard, Peiris, Tucker, Verde, Weiland, Wollack, \& Wright}]{page2007}
Page, L., Hinshaw, G., Komatsu, E., {et~al.} 2007, \apjs, 170, 335

\bibitem[{Planck-Collaboration(2005)}]{bluebook}
Planck-Collaboration. 2005, Planck: the Scientific Program, Vol.~1 (ESA-SCI)

\bibitem[{Planck-Collaboration(2011{\natexlab{a}})}]{early1}
Planck-Collaboration. 2011{\natexlab{a}}, A\& A accepted

\bibitem[{Planck-Collaboration(2011{\natexlab{b}})}]{early19}
Planck-Collaboration. 2011{\natexlab{b}}, A\&A submitted

\bibitem[{Planck-Collaboration(2011{\natexlab{c}})}]{early21}
Planck-Collaboration. 2011{\natexlab{c}}, A\&A submitted

\bibitem[{Planck-Collaboration(2011{\natexlab{d}})}]{early24}
Planck-Collaboration. 2011{\natexlab{d}}, A\&A submitted

\bibitem[{Planck-Collaboration(2011{\natexlab{e}})}]{early25}
Planck-Collaboration. 2011{\natexlab{e}}, A\&A submitted

\bibitem[{Planck-HFI-Core-Team(2010)}]{cote}
Planck-HFI-Core-Team. 2010, A\&A submitted

\bibitem[{Ponthieu {et~al.}(2005)Ponthieu, Mac\'ias-P{\'e}rez, \&
  Tristram}]{ponthieu2005}
Ponthieu, N., Mac\'ias-P{\'e}rez, J., \& Tristram, M. 2005, \aap, 444, 327

\bibitem[{Reich \& Reich(1988)}]{reich88}
Reich, P. \& Reich, W. 1988, AAS, 74, 7

\bibitem[{Reich {et~al.}(2004)Reich, Reich, \& Testori}]{reich04}
Reich, P., Reich, W., \& Testori, J. 2004, The magnetised interstellar medium,
  ed. B.~U. .~R. Wielebinski (Copernicus GmbH)

\bibitem[{Reid \& Brunthaler(2005)}]{reid}
Reid, M. \& Brunthaler, A. 2005, in ASP Conf. Ser. 340, Future Directions in
  High Resolution Astronomy: The 10th Anniversary of the VLBA, ed. J.~R. .~M.
  Reid (San Fransisco: ASP), 253

\bibitem[{Rybicki \& Lightman(1979)}]{ribicki}
Rybicki, G. \& Lightman, A. 1979, Radiative Process in Astrophysics (New York:
  Wiley-Interscience)

\bibitem[{Sun {et~al.}(2008)Sun, Reich, Waelkens, \& Ensslin}]{sun}
Sun, X., Reich, W., Waelkens, A., \& Ensslin, T. 2008, \aap, 477, 573

\bibitem[{Tauber {et~al.}(2010)Tauber, Mandolesi, Puget, \& Bouchet}]{tauber}
Tauber, J.~A., Mandolesi, N., Puget, J.~L., \& Bouchet, F. 2010, A\&A, 520

\bibitem[{Uyaniker {et~al.}(1999)Uyaniker, Furst, Reich, Reich, \&
  Wielebinski}]{uyaniker}
Uyaniker, B., Furst, E., Reich, W., Reich, P., \& Wielebinski, R. 1999, \aaps,
  138, 31

\bibitem[{Vaillancourt(2002)}]{vaillancourt}
Vaillancourt, J.~E. 2002, \apjs, 142, 335

\bibitem[{Wolleben {et~al.}(2006)Wolleben, Landecker, Reich, \&
  Wielebinski}]{wolleben}
Wolleben, M., Landecker, T., Reich, W., \& Wielebinski, R. 2006, \aap, 448, 411

\end{thebibliography}

\end{document}